# Semiconductor Nanopillar as a Single-Photon Emitter and its Optimal Design

S. Odashima[1, a)] and H. Sasakura[2, b)]

[1]*Research Institute for Electronic Science, Hokkaido University, Sapporo 001-0021, Japan*
[2]*Division of Applied Physics, Hokkaido University, Sapporo 060-8628, Japan*

a) Corresponding author: s_odashima@cris.hokudai.ac.jp
b) hirotaka@eng.hokudai.ac.jp

**Abstract.** The semiconductor quantum dot nanopillar array in InAs/GaAs was fabricated. In consideration of the quantum dot density, the pillar diameter was determined as only a few quantum dots were involved in a pillar. The lattice constant of a pillar array was optimized such that only one pillar couples to the fiber core without any specific manipulation. Such structural consideration regarding the photon source and the optical setup contributes to the high purity of the single-photon generation. For high extraction efficiency of photons from photon source, we fabricated the metal cavity structure around the pillar array. The combination of dry-etching and wet-etching realizes the functionally designed three-dimensional structure. Such fabrication processes face the realization of the well-optimized single-photon emitter.

## INTRODUCTION

Optical information technology based on photon number states provides highly secure quantum cryptography [1]. Among many requirements to the photon source, purity as a photon number state (single-photon, photon-pair, and so on), on-demand generation of photons, and long-term stability of photon output from the source are important factors to build up the fundamental researches to the actual technologies. Realizing such photon sources, semiconductor materials are useful because current semiconductor-processing techniques are available. In general, semiconductor mother materials for photon sources contain macroscopic number of photon emission points (quantum dots (QDs), impurities). To achieve well-defined photon number states, it is necessary to control the number of accessible photon emission points. The semiconductor QD nanopillar is appropriate for this purpose [2-4]. Considering the QD density of the mother sample, we can control the QD number involved in a pillar by changing the pillar diameter. In addition to the abovementioned matter, coupling between the photon source and optical equipment is also crucial. Fabrication of nanostructure on an optical fiber [2-10], cavity structure [11, 12], and DBR [13, 14] have been investigated for increasing the stability and efficiency of photon transmission.

In this study, we fabricate semiconductor nanopillars made by MBE-grown InAs/GaAs QDs sample. The pillar diameter and the lattice constant of an array are well considered so that only a few QDs couple to the measurement system. Furthermore, we fabricate metal cavity structures around pillars for better extraction efficiency of photons. To create such structures, we examine fabrication processes in detail. Manufactured samples show the functionally designed three-dimensional structure. Present semiconductor nanopillars and their structural design face to realizing the well-optimized single-photon emitter.

# EXPERIMENTS

The semiconductor InAs QDs were grown on a semi-insulating GaAs (001) substrate by molecular-beam-epitaxy (RIBER, MBE32P). After the growth of the GaAs buffer layer, InAs QDs were self-assembled, then covered by a GaAs cap layer of 50 nm. On the top surface, QDs were regrown to investigate the growth condition of QDs. For well-defined photon number states, the accessible QD number must be restricted. Therefore, we fabricated the nanopillar array for artificially reducing the number of QDs which couple to the optical path for measurement, by using the procedure shown below. The pillar diameter was determined considering the number of QDs involved in a pillar. In addition to the pillar array structure, we also fabricated the metal cavity around nanopillars by a combination of dry-etching, wet-etching, and metal deposition. The fabrication processes will be discussed in detail in the following sections.

## Optimization of a Pillar Array for Single-Photon Coupling to a Fiber

The pillar array structure was fabricated by EB lithography (ELIONIX, ELS-F125-U) with HSQ (Dow Corning Toray, Fox(R) 15 Flowable oxide), and reactive ion etching (SAMCO, RIE-101iHS) with $Cl_2$/Ar gases under the pressure of 0.08 Pa. The bias and ICP voltage of RIE were 160 W and 32 W, respectively. The low pressure and high ratio of bias/ICP enabled anisotropic etching, then we could get the straight-shaped pillar. The pillar diameter is about 300 nm. The pillar array has a square lattice structure with a lattice constant of 2.5 µm. Figure 1 shows SEM images of such a prepared sample. We can observe 5, 6 QDs on the top surface of a pillar. The estimated QD density is about $7\times10^9$ /cm$^2$. It is worth mentioning that QDs located near the edge of a pillar are optically inactive because of the weakness of confinement to QDs, and/or mechanical and chemical damages by etching. Therefore, only one or two QDs near the center of a pillar contribute to photon emission. Accessing to QDs in a pillar, we use the single-mode fiber with a mode diameter of 2.6 µm at 1,100 nm (Thorlabs, UHNA3). Therefore, if this pillar array sample directly contacts the fiber end-surface, only one pillar can couple to the fiber without any specific manipulation [4]. Such geometrical consideration is important to avoid ambiguity of the coupled QD number to the measurement system.

## Metal Cavity Structure

We fabricated metal cavity structure which covers a semiconductor QD pillar array, expecting high extraction efficiency of photons. First, we fabricated pillars which have a diameter of about 1.2 µm (Fig. 2(a)). The height, except for the HSQ mask is about 280 nm. This pillar sample was etched with 10% $H_2SO_4$ solution with the additive of $H_2O_2$. This wet-etching is isotropic to GaAs. Therefore, the GaAs pillar part becomes thin, and the bottom part becomes gradually wider, merging with the base plane. The HSQ mask, on the other hand, is tolerant of this wet-etching. The mask keeps the original size during the wet-etching, then we have the mushroom-like shape as shown in Fig. 2(b). After the wet-etching, the height of the InAs/GaAs semiconductor pillar became about 550 nm. Considering the height of this semiconductor pillar, Ti of 10 nm and following Au of 540 nm were deposited by EB evaporation (EIKO, EB-580S). After removing the HSQ mask by 8 mol/L KOH solution, we had semiconductor pillars covered by Au/Ti metal cavity (Fig. 2(c)). As shown in Fig. 2(c), there is a space between the sidewall of a semiconductor pillar and a metal cavity. This space prevents the excited carrier in QD from diffusing into the metal side. The size of the HSQ mask is directly reflected in the size of the metal aperture, and the space between the pillar and metal is controlled by semiconductor wet-etching time. Figure 3 shows the wet-etching time dependence of a pillar diameter (a) and a pillar height (b). The etchant is 10% $H_2SO_4$ solution and the additive of 30% $H_2O_2$ with the ratio of 400: 1. To estimate the wet-etching rate, we prepared GaAs pillars with diameters of 800, 900, 1000, 1,100, and 1,200 nm (CAD size of EB lithography). SEM observation evaluated each pillar diameters by changing the etching time (Fig. 3(a)). Our results show the etching rate of 9.74 nm/min. Results of pillar height also support this estimation (Fig. 3(b)). By combining dry-etching and wet-etching, and the following metal deposition procedure, we can fabricate the functional three-dimensional structure suitable for a resonance of photons generated from QD.

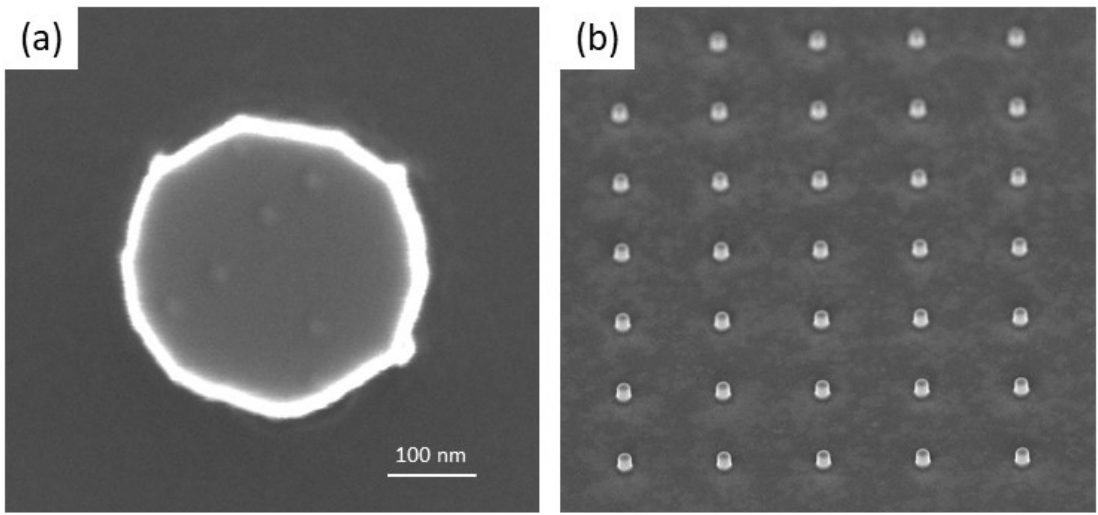

**FIGURE 1.** (a) SEM image of a pillar. QDs on the top surface are for investigating QD growth condition. (b) Square lattice pillar array structure with a lattice constant of 2.5 µm.

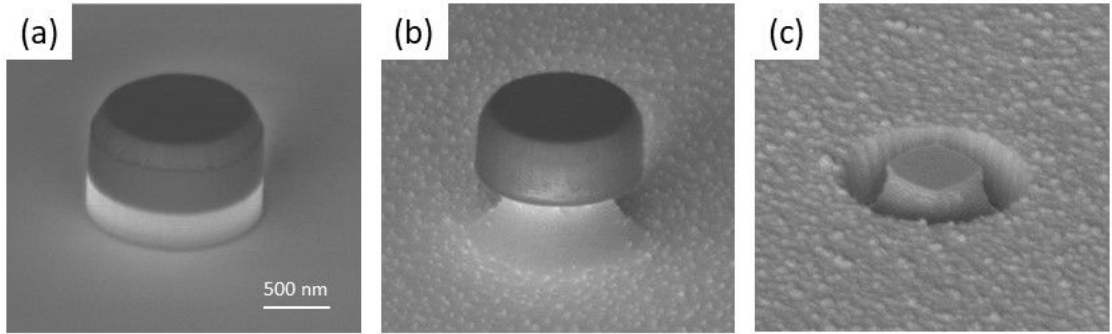

**FIGURE 2.** (a) InAs/GaAs QD semiconductor pillar with HSQ mask. (b) After wet-etching by 10% $H_2SO_4$ with $H_2O_2$. Semiconductor part is isotropically etched. (c) After Au/Ti deposition, HSQ mask is removed. Semiconductor pillar is covered by Au/Ti.

For photoluminescence (PL) measurements of this sample, we used a He-Ne laser for excitation. A double-grating spectrometer (Action, Spectrapro 2500i, $f$ = 1.0 m) and a liquid-nitrogen-cooled InGaAs photodiode array (Roper, Pyron-IR) were used to evaluate PL spectra. The PL spectrum of an InAs/GaAs semiconductor pillar with a metal cavity is shown in Fig. 4(a). This result shows the well-defined emission peak at 957.5 nm. Our approach to control the QD number in a pillar leads to discrete peaks, which makes it possible to identify the contribution of each QD. On the emission peak at 957.5 nm, we performed an auto-correlation measurement using optical fiber based Hanbury Brown and Twiss interferometer equipped with a pair of superconducting nanowire single-photon detectors (Single Quantum, custom-made product). The coincidence counts were recorded by a time-correlated single-photon-counting module (Becker & Hickl, SPC-130EM). The second-order correlation function, $g^{(2)}(t)$ is the normalized coincidence with a relative time $t$ (Fig. 4(b)). Therefore, at $t = 0$ value, $g^{(2)}(0)$ indicates purity as a single-photon emission. Our result, $g^{(2)}(0) = 0.195$ of the discrete emission peak, suggests that this InAs/GaAs QD semiconductor nanopillar with a metal cavity has the potential as a functionally designed single-photon emitter. Further investigation will be performed in the future.

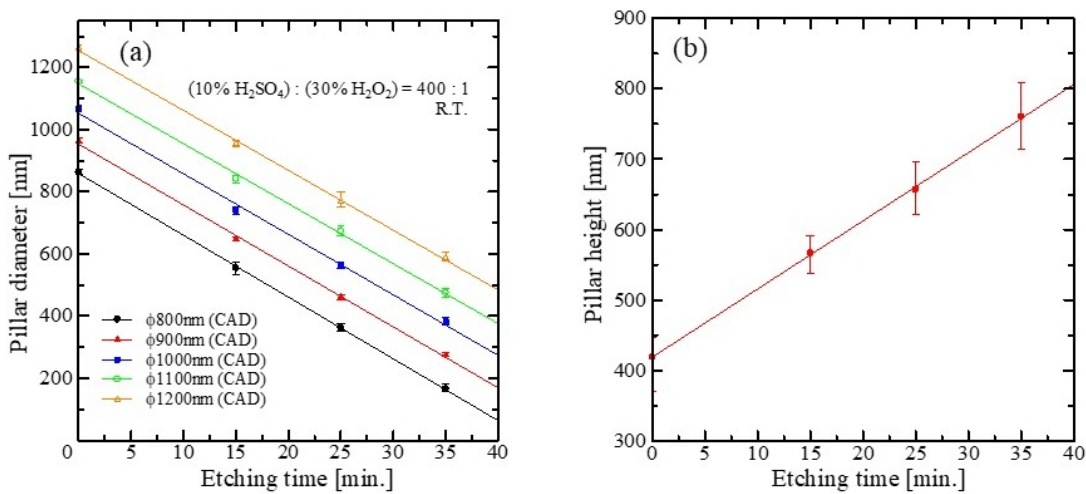

**FIGURE 3.** Wet-etching time dependence of the pillar diameter (a), and pillar height (b). The estimated etching rate is 9.74 nm/min.

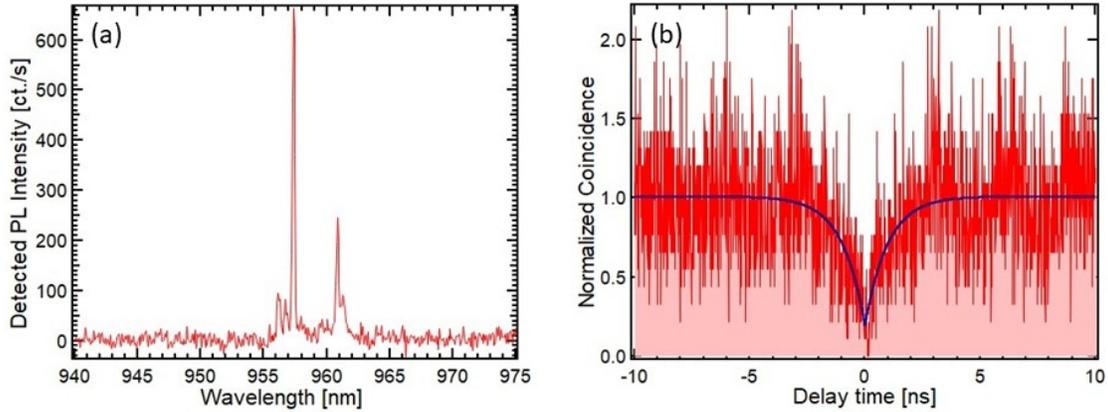

**FIGURE 4.** (a) PL spectrum of the InAs/GaAs QD nanopillar covered by Au/Ti metal cavity. (b) Auto-correlation measurement of 957.5 nm emission line. $g^{(2)}(0)$ is 0.195.

## CONCLUSION

Semiconductor InAs/GaAs QD nanopillar array was fabricated. The pillar diameter was optimized based on the QD density, as only a few QDs contribute to the photon generation in a manner of the well-defined single-photon emission. The metal cavity structure around pillars was fabricated by a combination of dry-etching and wet-etching, and the following metal deposition. We can control the cavity size by adjusting the dry-etching mask size and the wet-etching time. Such structural design and nanoprocessing techniques realize the well-optimized single-photon emitter.

## ACKNOWLEDGMENTS

This work is supported by JSPS KAKENHI Grant No. 20H0255500. Samples were fabricated by using equipment of nanotechnology platform, Hokkaido University.